\documentclass[a4paper,10pt]{amsart}
\usepackage{xcolor}
\usepackage[utf8]{inputenc}
\usepackage[a4paper, left=4.0cm, right=4.0cm, top=3.0cm, bottom=3.5cm, includefoot]{geometry}
\usepackage[
    hyperref, backend=biber, doi=false,url=false,sorting=none,backref=true]{biblatex}
\DefineBibliographyStrings{spanish}{andothers={\textsl{et al}\adddot}}
\renewbibmacro{in:}{}

\usepackage{amsmath, amssymb,stackrel,amsaddr}
\usepackage{hyperref}
\usepackage[english]{babel}
\usepackage{graphicx}

\hypersetup{colorlinks=true,linktoc=page,linkcolor=blue,citecolor=red,urlcolor=cyan}

\bibliography{bibliografia}

\title{Snyder-de Sitter meets the Grosse-Wulkenhaar model}

\author{S.~A.~Franchino-Vi\~nas}
\address{Departamento de F\'isica, Facultad de Ciencias Exactas
Universidad Nacional de La Plata, C.C.\ 67 (1900), La Plata, Argentina,}
\address{Theoretisch-Physikalisches Institut, Friedrich Schiller Universit\"at Jena, Max Wien Platz 1, 07743 Jena, Germany.}
\email{sa.franchino@uni-jena.de}
\author{S. Mignemi}
\address{Dipartimento di Matematica e Informatica, Università di Cagliari, viale Merello 92, \\09123 Cagliari, Italy,}
\address{INFN, Sezione di Cagliari, Cittadella Universitaria, 09042 Monserrato, Italy.}
\email{smignemi@unica.it}

\begin{document}

\begin{abstract}
{We study an interacting $\lambda\,\phi^4_{\star}$ scalar field defined on Snyder-de Sitter space. Due to the {noncommutativity} as well as the curvature of this space, the renormalization of the two-point function differs from the commutative case. In particular, we show that the theory in the limit of small curvature and noncommutativity is described by a model similar to the Grosse-Wulkenhaar one. Moreover, very much akin to what happens in the Grosse-Wulkenhaar model, our computation demonstrates that there exists a fixed point in the renormalization group flow of the harmonic and mass terms. }
 
\end{abstract}

\maketitle

\section{Introduction}
\label{sec:introduction}
The interplay between physics and mathematics is well-known. One of the chapters of its history began in the 1940's, as Snyder introduced what we nowadays call a noncommutative (NC) space as a way to avoid the infinities that arise in Quantum Field Theory (QFT) \cite{Snyder:1946qz}. The underlying idea of his work was to introduce a high-energy cut-off by means of a discretization of space. Although obscured by the success of renormalization techniques, noncommutative theories were revived in the late 1980's by a series of both formal and applied papers by Connes and collaborators on spectral actions \cite{Connes:1990qp,Connes:1994yd}.

The number of publications increased considerably in the following years, cf. \cite{Landi:1997sh, Douglas:2001ba} for a review. The main reason for such a success, besides its mathematical beauty, was the exhibition of many expected features of an effective Quantum Theory of Gravity. However, among the most resonant results was the UV/IR mixing occurring in some of these theories \cite{Seiberg:1999vs}. This, in conjunction with the fact that many of these theories were built in  scenarios of broken Lorentz invariance, entailed many detractors to NC QFT. However, the previously mentioned facts are not inherent properties of NC QFT.

Indeed, a great impact was caused by the Grosse-Wulkenhaar model \cite{Grosse:2004yu}, a scalar field model which remedied not only the UV/IR mixing problem but showed many other interesting properties regarding constructiveness, solvability and integrability \cite{Grosse:2013lln,Grosse:2018npj}. On the other hand, two of the most interesting solutions to the Lorentz non-invariance problem are the deformation (or twisting) of the symmetry algebra \cite{Chaichian:2004za} (in the so called Moyal plane) or the consideration of the original Snyder space. In this regard several developments have been achieved in the last years -- among others, the behaviour of particles in Snyder space \cite{Lu:2011fh,Mignemi:2011gr,Mignemi:2013aua}, the star product, the coproduct and antipodes of its Hopf algebra, the twist operator \cite{Battisti:2010sr,Meljanac:2009fy,Meljanac:2011mt,Meljanac:2017qck}, and a scalar QFT defined on it \cite{Meljanac:2017ikx,Meljanac:2017grw,Franchino-Vinas:2018jcs} have been studied.

In this context a deformation of Snyder spaces in the spirit of triple special relativity \cite{KowalskiGlikman:2004kp}, the Snyder-de Sitter space (SdS), has also been pursued \cite{Mignemi:2011wh,Mignemi:2009zz,Ivetic:2013yga,Mignemi:2015una}. Nevertheless, the similarities between SdS and the Grosse-Wulkenhaar model have apparently passed unnoticed -- to revert this situation is the goal of this work, which is organized as follows: in Section \ref{sec:SdS} we review the basics of SdS model. Then, we expose the similarities between SdS and the Grosse-Wulkenhaar model in Section \ref{sec:2point}. Finally we draw our conclusions in Section \ref{sec:conclusions}.

\section{Snyder-de Sitter model}
\label{sec:SdS}
The Snyder-de Sitter space (SdS) is a generalization of the Snyder model to curved spacetime. We will deal with its Euclidean D-dimensional version, defined by the following algebra of position ($\hat{x}_i$) and momentum ($\hat{p}_j$) operators
\begin{align}\label{eq:commutation}
 &[\hat{x}_i,\hat{x}_j]=i\beta^2 J_{ij},\qquad [\hat{p}_i,\hat{p}_j]=i\alpha^2 J_{ij},\\
 &[\hat{x}_i,\hat{p}_j]=i[\delta_{ij}+\alpha^2 \hat{x}_i\hat{x}_j+\beta^2\hat{p_j}\hat{p}_i+\alpha\beta (\hat{x}_j\hat{p}_i+\hat{p_i}\hat{x}_j)],
\end{align}
where $i,j=1,\cdots,D$, and $J_{ij}=\frac{1}{2}(x_ip_j-x_jp_i+p_jx_i-p_ix_j)$ are the (hermitian) generators of the Poincar\'e algebra, defined in terms of position and momentum operators ($x$ and $p$) satisfying canonical commutation relations (see below). Notice that the action of these generators on position and momentum space is undeformed.

A relevant property of the SdS model is that it can be rewritten by means of a noncanonical transformation into the usual Snyder space -- indeed, the operators $X_i$ and $P_i$ defined by \cite{Mignemi:2009zz,Mignemi:2015una}
\begin{align}\label{eq:2snyder}
 \hat{x}_i=:X_i, &\qquad \hat{p}_i=:P_i -\frac{\alpha}{\beta} X_i,
\end{align}
are the Snyder operators, i.e. they satisfy the commutation relations
\begin{align}\label{Sny}
 [X_i,X_j]=i\beta^2 J_{ij},& \qquad [P_i,P_j]=0,&\qquad [X_i,P_j]=i(\delta_{ij}+\beta^2P_iP_j).
\end{align}
Of course, (\ref{Sny}) can be seen as the flat space limit ($\alpha\rightarrow 0$) of eq. \eqref{eq:commutation}. However, it should be noticed that the noncanonical transformation \eqref{eq:2snyder} is singular for vanishing $\beta$. In other words, what would be called the de Sitter limit cannot be recovered.

In turn, the Snyder algebra can be expressed in terms of canonical operators ($x_i$, $p_i$), i.e. in terms of derivative operators, as
\begin{align}\label{eq:2canonical}
 P_i=:p_i=-i\partial_i, \quad X_i=:x_i+\beta^2x_jp_jp_i=x_i-\beta^2x_j\partial_j\partial_i.
\end{align}
At this point the non-hermiticity of the operator $X_i$ is clear. Nonetheless, it can be made Hermitian by performing a symmetrization very much akin the one employed in \cite{Meljanac:2017ikx},
\begin{align}\label{eq:symmetrization}
 X_i\rightarrow X_i=x_i+\frac{\beta^2}{2} (x_jp_jp_i+p_ip_jx_j).
\end{align}
In particular, after inserting eqs. \eqref{eq:2canonical} and \eqref{eq:symmetrization} in \eqref{eq:2snyder}, it follows that in this realization the SdS momentum operator can be written as
\begin{align}\label{eq:phat}
 \hat{p}_i=p_i-\frac{\alpha}{\beta} x_i-\frac{\alpha \beta}{2} (x_jp_jp_i+p_ip_jx_j).
\end{align}

Before we come to the definition of a scalar QFT on SdS, still another ingredient will prove useful. As is usual in NC QFT, the commutator of the position operators can be alternatively described by a deformed product, called star product. For this realization the star product $\star$ can be found to be\footnote{Notice our change in the conventions with respect to \cite{Meljanac:2017ikx}, $\beta\rightarrow \beta^2$.} \cite{Meljanac:2017ikx}
\begin{align}
 e^{ik\cdot x}\star e^{i q \cdot x}=\frac{e^{iD(k, q) x}}{(1-\beta^2 k\cdot q)^{5/2}},
\end{align}
where
\begin{align}
 D_{\mu}(k,q):=\frac{1}{1-\beta^2 k \cdot q} \left[\left(1- \frac{\beta^2 k \cdot q}{1+\sqrt{1+\beta^2 k^2}}\right)k_{\mu}+ \sqrt{1+\beta^2 k^2} q_{\mu}\right].
\end{align}

Now an interacting scalar QFT can be defined by introducing an action functional. We will consider in this article the simplest case where the kinetic term is given by the momentum operator in SdS space, $\frac{1}{2}(\hat{p}^2+m^2)$, and the interaction is given by a quartic potential. Actually, an SdS-invariant kinetic term would be more involved, but the final result would contain essentially the same kind of contributions. In formulae, we consider the action
\begin{align}\label{eq:action}
S=\frac{1}{2} \int d^Dx\; \varphi \left( \hat{p}^2+m^2\right) \varphi -\frac{\lambda}{4!} \int d^Dx (\varphi \star \varphi)\star (\varphi \star \varphi).
\end{align}
Then, after a straightforward expansion of $\hat{p}_i$ in terms of the canonical operators we can recast the quadratic terms in the action as
\begin{align}
S^{(2)}=\frac{1}{2} \int d^Dx\; \varphi A \varphi,
\end{align}
where the operator $A$ is defined as
\begin{align}
 \begin{split}
A &:= \left( 1+i \frac{(D+1)}{2}\alpha \beta\right)^2 p^2+\frac{\alpha^2}{\beta^2}x^2+m^2+ \beta^2\alpha^2 x_j p_j p_i x_kp_kp_i \\
 &-\beta \alpha \left(1+i \frac{D+1}{2}\alpha \beta\right) (- p\cdot p p\cdot x
 + p_i x \cdot p p_i )
 +\alpha^2 (x_ix_jp_jp_i+x_jp_jp_i x_i),
 \end{split}
 \end{align}
 and boundary terms have been discarded.
 This expression can be further simplified by taking into account the physical assumption that both curvature and noncommutativity should be small. As a consequence we can ignore higher order terms in $\alpha$ and $\beta$, and hence
 \begin{align}
 A= p^2+\frac{\alpha^2}{\beta^2}x^2
 +\alpha^2 (x_ix_jp_jp_i+x_jp_jp_i x_i)+ \mathcal{O}(\gamma^4).
\end{align}
where $\gamma$ is the defining scale of both $\alpha$ and $\beta$.
We are thus left with a Grosse-Wulkenhaar-like action to which some small corrections are added. The fact that the harmonic oscillator term has a pure geometric origin is rather striking, even if this is not the first time that a link between the harmonic term and the geometry is proposed \cite{Buric:2009ss}. It is therefore a natural question to ask ourselves whether this model exhibits properties similar to those of the Grosse-Wulkenhaar one or not. In the next section, we shall address the two-point function renormalization.

\section{Two-point function renormalization}\label{sec:2point}
Once we have obtained expression \eqref{eq:action} for the action, we can perform a perturbative computation in order to analyze the loop corrections to the effective action. This can be done using the well-known Feynman diagram's expansion-- there exists however a specially well-suited technique called the Worldline or String-inspired formalism \cite{Schubert:2001he}. Some previous implementations of this technique in noncommutative theories may be found in \cite{Bonezzi:2012vr,Vinas:2014exa,Franchino-Vinas:2015vpa,Franchino-Vinas:2018jcs}, whereas details in the computation of the two-point function result that concerns us can be read in \cite{Franchino-Vinas:2019}. In particular, the one-loop divergent part of the two-point {function in $D=4-\varepsilon$ dimensions, $\Gamma^{(2)}_{div}$, is
 \begin{align}
\begin{split}
\Gamma^{(2)}_{div}[\phi]&= \frac{\lambda}{192\pi^2\varepsilon} \int d^4x\, \phi(x) \Bigg[ -4\omega^2  \beta ^2  \left(2 x_{\mu}x_{\nu} \partial^{\mu}\partial^{\nu} +x^2 \partial^2 \right)\\\
&\hspace{-0.5cm} +x^2 \left(15 \alpha_{\text{eff}}^2 m^2 + 6 \omega^2 + 36 m^2 \beta^2 \omega^2\right)+x^4 \left(9 \alpha_{\text{eff}}^2 \omega^2 + 24 \beta^2 \omega^4\right)\\
 &\hspace{2.5cm} -26 \alpha_{\text{eff}}^2 + 6 m^2 + 12 m^4 \beta^2 + \frac{6 \alpha_{\text{eff}}^2 m^4}{\omega^2} + 48 \beta^2 \omega^2\Bigg] \phi(x),
\end{split}
 \end{align}
where we have defined $\omega:=\frac{\alpha}{\beta}$} and $\alpha_{\text{eff}}=2\alpha$. From this expression, one can readily observe some similarities and differences with respect to the Grosse-Wulkenhaar case.

On one side, there are additional harmonic terms proportional to quadratic differential operators. These should be traced to the way the star product in Snyder spaces modifies the quartic potential and their appearance is no surprise. Indeed, this kind of terms can in principle be interpreted as coming from an effective metric in configuration space \cite{Franchino-Vinas:2018jcs} -- to be concrete consider for example the metric $g_{\mu\nu}$ of de Sitter space in projective coordinates \cite{Mignemi:2015una}, whose determinant reads $g=(1+\alpha^2x^2)^{D+1}$. 

On the other side, also the harmonic term and the mass should be renormalized. In this one-loop approximation, the former possesses a fixed point corresponding to a curvature parameter $\alpha$ that solves a quadratic equation for a given mass and noncommutative parameter $\beta$. At this point also the renormalization of the mass gets simplified and in order to find the fixed points, an equation must be solved for the noncommutativity parameter $\beta$. Although these clues are of course not decisive, they could point to a general property of curved noncommutative spaces regarding fixed points.

Finally, one must take into account that an exact treatment of the dependence on $\alpha$ and $\beta$, rather than a series expansion, should improve the divergences of the theory.

\section{Conclusions}\label{sec:conclusions}
The study of an interacting scalar QFT defined on SdS shows many stimulating features. First of all it confirms the idea, presented in \cite{Buric:2009ss} and developed in successive works, that the harmonic term present in the Grosse-Wulkenhaar model could be linked to a geometrical aspect of the theory. In particular, and in the spirit of the triply special relativity, this geometrical aspect is related to some interaction between the noncommutativity of the space and its curvature, ultimately linked to the cosmological constant.

Moreover, the two-point function of our scalar theory displays divergences that imply the renormalization of several terms. In spite of what could be guessed at a first sight, this doesn't purport a proliferation of the to-be-renormalized terms. In effect, the only additional terms in comparison to the Grosse-Wulkenhaar model are those proportional to a quadratic differential operator, and could be therefore ascribed to the appearance of an effective metric.

In turn, the renormalization of the harmonic term exhibits a fixed point that resembles that observed in the Grosse-Wulkenhaar case. However, the former entails a relation between the curvature, the noncommutative parameter and the mass of the field, whereas the latter implies a link between the noncommutative parameter and the frequency of the oscillator. This could be a hint that some other features of the Grosse-Wulkenhaar model could be present in SdS or even in a more general class of models in curved noncommutative spaces.
It would be also of interest to analize whether the model possesses a hidden symmetry or satisfies a kind of Ward identity, such as in the case of the Grosse-Wulkenhaar model, and their connection to the renormalization properties of the model.

Needless to say, a deeper analysis taking into account the exact de Sitter kinetic term and analyzing the renormalization of the remaining $n$-point functions (especially the four-point function\cite{Franchino-Vinas:2019}) should be performed in order to elucidate these aspects.

\section{Acknowledgement}
This work was partially supported by GESTA - Fondazione di Sardegna. SAF acknowledges support from  the DAAD and the Ministerio de Educaci\'on Argentino under the ALE-ARG program. SAF would like to thank the Universit\`a di Cagliari and specially  SM for their  hospitality.

\printbibliography

\end{document}